\documentclass[a4paper,11pt]{article}
\usepackage{pos}
\usepackage{amsmath,graphicx}
\usepackage[english]{babel}
\usepackage[utf8]{inputenc}  
\usepackage[T1]{fontenc}
\usepackage{slashed} 
\usepackage{ifthen}
\newcommand{\dInt}[2][]{%
    \ifthenelse{\equal{#1}{}}
    {\ensuremath{\operatorname{d}{#2}\;}}
    {\ensuremath{\operatorname{d}^{#1}{#2}\;}}
}

\newcommand{\Proj}[1]{{P_\text{#1}}}
\def\La{{\cal L}}

\title{$\gamma_5$ in dimensional regularization --- a no-compromise approach using the BMHV scheme}
\ShortTitle{$\gamma_5$ in dimensional regularization using the BMHV scheme}

\author[a]{Hermès Bélusca-Maïto}
\author[a]{Amon Ilakovac}
\author[a]{Marija Mađor-Božinović}
\author[b]{Paul Kühler}
\author*[b]{Dominik Stöckinger}

\affiliation[a]{Department of Physics, University of Zagreb,\\ Bijeni\v{c}ka cesta 32, HR-10000 Zagreb, Croatia}

\affiliation[b]{Institut für Kern- und Teilchenphysik, TU Dresden,\\ Zellescher Weg 19, DE-01069 Dresden, Germany}

\emailAdd{dominik.stoeckinger@tu-dresden.de}

\abstract{$\gamma_5$ is notoriously difficult to define in $D$ dimensions. The traditional BMHV scheme employs a non-anticommuting $\gamma_5$. Its key advantage is mathematical consistency and the existence of all-order proofs. Its disadvantage is the spurious breaking of gauge invariance in chiral gauge theories like the electroweak standard model. Our research programme aims to determine the special finite counterterms which are necessary to restore gauge invariance, to allow more straightforward applications of the BMHV scheme and to cross-check alternative schemes. In these proceedings we present the key concepts and methods, and we outline the calculational procedure and present results for an abelian gauge theory at the 2-loop level. An important observation is the simplicity of the results --- three types of symmetry-restoring counterterms are sufficient at the 2-loop level. }

\FullConference{%
  Loops and Legs in Quantum Field Theory - LL2022,\\
  25-30 April, 2022\\
  Ettal, Germany
}

\begin{document}
\maketitle

\section{Introduction}

The problem of $\gamma_5$ in dimensional regularization is well
known. The three properties $(i)$ anticommutativity with $\gamma^\mu$,
$(ii)$ non-zero
$\text{Tr}(\gamma_5\gamma^\mu\gamma^\nu\gamma^\rho\gamma^\sigma)$,
$(iii)$ cyclicity of traces become inconsistent in $D\ne4$ dimensions
[e.g.\ $(i)$ and $(iii)$ imply that the trace in $(ii)$ is zero].

There is a multitude of proposals how to treat this issue and how to
define a $D$-dimensional continuation of $\gamma_5$ and many of them
are routinely applied in practical computations (for a review see
\cite{Jegerlehner:2000dz}, for further original references see also
\cite{Belusca-Maito:2020ala,Belusca-Maito:2021lnk}).

A very traditional scheme is the original proposal of
Ref.\ \cite{tHooft:1972tcz}, which was later further formalized in
Ref.\ \cite{Breitenlohner:1975hg} --- the BMHV scheme. It is well
known that this scheme has significant disadvantages in practical
calculations. However, its key advantage is that full mathematical
consistency and complete all-order proofs are established
\cite{Breitenlohner:1975hg}. In our approach we aim to avoid
compromises with respect to mathematical rigor. Hence this is our
motivation to focus on the BMHV scheme. We accept its practical
difficulties, deal with them, and aim to provide the community with
results and building blocks which allow more straightforward
applications of the scheme.\footnote{%
  In addition, a better understanding of the BMHV scheme may feed back
to alternative approaches to $\gamma_5$, potentially enabling
consistency checks or optimizations of such approaches.}

Specifically we aim to provide the
required symmetry-restoring counterterms which compensate the spurious
breaking of gauge invariance caused by the non-anticommuting
$\gamma_5$.
Ultimately we aim for a treatment of the electroweak standard model at
the multiloop level. The current status is a treatment of a general
Yang-Mills theory at the 1-loop level \cite{Belusca-Maito:2020ala} 
and an abelian gauge theory at
the 2-loop level \cite{Belusca-Maito:2021lnk}.
In these proceedings we provide an introduction to the key concepts
and methods of our approach (sec.\ 2 and 3) and an outline of the
computations and results for the abelian case (sec.\ 4). Sec.\ 5
contains a brief summary and outlook.

\section{Definitions and the problem in a nutshell: breaking of Ward identity}
\label{nutshell}
In the BMHV scheme,  formally $D=(4-2\epsilon)$-dimensional quantities $k^\mu$ of
dimensional regularization can be split into their $4$-dimensional and
$(D-4)$-dimensional parts as
\begin{align}
  k^\mu &= \bar{k}^\mu + \hat{k}^\mu \,.
\end{align}
The $D$-dimensional space can be viewed as a direct sum of
$4$-dimensional and $(D-4)$-dimensional subspaces, such that
orthogonality and projection relations such as
\begin{align}
  \bar{k}^\mu \hat{k}_\mu & = 0 \,,&
  k^\mu \hat{k}_\mu & =   \hat{k}^\mu \hat{k}_\mu \,,&
  k^\mu \bar{k}_\mu & =   \bar{k}^\mu \bar{k}_\mu 
\end{align}
hold. The split can be done for objects such as momentum vectors,
gauge fields, metric tensors, and in particular for $\gamma^\mu$
matrices, $\gamma^\mu = \bar{\gamma}^\mu+\hat{\gamma}^\mu$.

In the BMHV scheme, the matrix $\gamma_5$  is defined as an
intrinsically 4-dimensional object. It satisfies
  \begin{align}
\label{eq:Gamma5DReg_A}
    \{\gamma_5, \bar{\gamma}^\mu\} &= 0 \, , &
    [\gamma_5, \hat{\gamma}^\mu] &= 0 \, ,
  \end{align}
and thus it breaks full $D$-dimensional Lorentz covariance. The
usual anticommutation relation holds only for the purely $4$-dimensional
parts of the $\gamma^\mu$ matrices. Importantly, this definition is
consistent with 
the cyclicity of traces and with the relation $\gamma_5 = \frac{-i}{4!}
\epsilon_{\mu\nu\rho\sigma} \bar{\gamma}^\mu \bar{\gamma}^\nu \bar{\gamma}^\rho
\bar{\gamma}^\sigma$.

\begin{figure}[t]
{    \includegraphics[width=.9\textwidth]{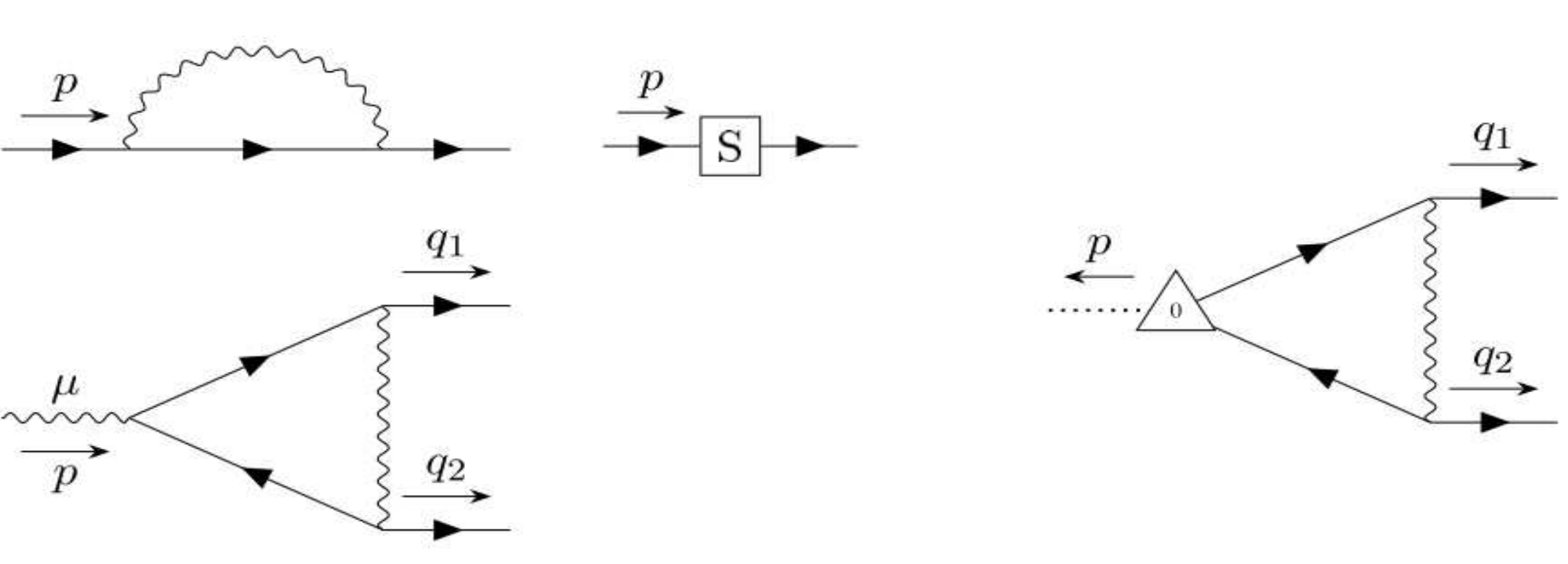}
}
    \\
    \null\hspace{.13\textwidth}(a)\hspace{.24\textwidth}(b)\hspace{.3\textwidth}(c)
\caption{\label{fig:WI}Illustration of (a) diagrams which violate a
  Ward identity, (b) a counterterm restoring this Ward identity, (c) a
  single diagram with the insertion of an operator $\widehat\Delta$
  which directly yields the breaking of the Ward identity (a).}
\end{figure}

Let us provide a preview of the main problem caused by the definition
(\ref{eq:Gamma5DReg_A}), the breaking of gauge invariance in chiral
gauge theories. In the abelian gauge theory defined below we expect the validity of
QED-like Ward identities such as a relationship between the one-loop
fermion self energy and the one-loop fermion--gauge boson three-point
function, as illustrated in Fig.\ \ref{fig:WI}(a).

It turns out that the corresponding Ward identity is violated at the
level of the BMHV regularized one-loop diagrams. The breaking has a
special form, however: it is local, i.e.\ it has a form which can be
compensated by adding a certain local, symmetry-restoring counterterm
to the Lagrangian. The counterterm contributes only to the fermion
self energy as illustrated in Fig.\ \ref{fig:WI}(b).
After adding the counterterm, there is an
additional contribution to the fermion self energy, and the Ward
identity is fulfilled.

The precise form of the counterterm action relevant here and
corresponding to the \boxed{S} in Fig.\ \ref{fig:WI}(b) is
\begin{align}
\begin{split}
	S^1_\text{fct} =
	\frac{e^2}{16\pi^2} \int \dInt[4]{x} &\Bigg\{
\ldots	+ \left(  \frac{5+\xi}{6} \right) (\mathcal{Y}_R^j)^2 			\Big( \bar{\psi}_j \dot\imath \bar{\slashed{\partial}} \,\Proj{R}\, \psi_j \Big)
\Bigg\}
	\, .
\end{split}
\label{eq:FiniteCT1Loop}
\end{align}
The problem of the BMHV scheme is thus that Ward/Slavnov-Taylor
identities are broken in intermediate steps and special,
symmetry-restoring counterterms are needed. The existence of such
counterterms is guaranteed provided the gauge theory in question is
free of chiral anomalies and hence renormalizable. But their concrete
determination is a necessary step and a complication of practical
computations.

\section{Goals and method: Slavnov-Taylor identities and quantum action principle}

In a nutshell, our goal is to determine symmetry-restoring
counterterms such as Eq.\ (\ref{eq:FiniteCT1Loop}), for all Ward and
Slavnov-Taylor identities, and at the multi-loop level.

In principle, a pedestrian way to do that might be to evaluate all
Green functions entering relevant Ward and Slavnov-Taylor identities,
check the validity of the identities, and ultimately evaluate possible
breakings and required counterterms. There is, however, a more direct
method, which can also be illustrated with the example introduced
above.\footnote{%
  For more details on both methods and literature references with
  sample applications see sec.\ 6 of \cite{Belusca-Maito:2020ala}.}
Instead of evaluating the fermion self-energy and
fermion--gauge boson three-point function, it is sufficient to
evaluate the Feynman diagram in Fig.\ \ref{fig:WI}(c).

In this diagram, the triangle denotes the insertion of a special
operator $\widehat{\Delta}$, which can be determined once and for all as
will be described below. The quantum action principle guarantees that the result of this diagram corresponds directly to
the {\em violation} of the Ward identity of Fig.\ \ref{fig:WI}(a). Hence, in order to
compute the required counterterm (\ref{eq:FiniteCT1Loop}) we only need
to compute the single diagram  Fig.\ \ref{fig:WI}(c) instead of the two diagrams
 Fig.\ \ref{fig:WI}(a). In addition, the single diagram Fig.\ \ref{fig:WI}(c) is simpler to
compute since the operator $\widehat{\Delta}$ is evanescent, i.e.\ zero in 4
dimensions, and therefore only terms related to ultraviolet $1/(D-4)$
singularities can lead to non-vanishing contributions.

In general, our method is therefore to compute all relevant breakings
of Ward/Slavnov-Taylor identities in terms of such Feynman diagrams
based on the quantum action principle, and then to determine the
required symmetry-restoring counterterms.\footnote{%
  We remark that the method of computing potential symmetry breakings
  based on the quantum action principle was also used in
  Refs.\ \cite{Stockinger:2005gx,Hollik:2005nn,Stockinger:2018oxe} in the study of SUSY properties of dimensional
  reduction. In those references the method established that
  dimensional reduction {\em preserves} SUSY in important cases up to
  the 3-loop level --- the corresponding diagrams involving the
  insertion $\widehat{\Delta}$ turned out to vanish.}

\section{Application to abelian chiral gauge theory at the 2-loop level}

In this section we outline the concrete calculational procedure and the
results for the 2-loop renormalization of chiral gauge theories in the BMHV
scheme. We focus on an abelian gauge theory similar to the U(1)
hypercharge sector of the electroweak standard model.
The essential steps are:

\begin{minipage}{.9\textwidth}  
\begin{flushleft}
\begin{description}
\item 1.\ Define $D$-dimensional regularized Lagrangian and compute
  the resulting symmetry
  breaking $\widehat{\Delta}$.
\item 2.\ Determine 1-loop UV divergences and the resulting
  counterterm Lagrangian $\La_{\text{sct}}$.
\item 3.\ Determine 1-loop violation of Slavnov-Taylor
  identity using the quantum action principle and the insertion $\widehat{\Delta}$.
\item 4.\ Determine 1-loop symmetry-restoring counterterms and the
  resulting, finite, symmetry-restoring counterterms $\La_{\text{fct}}$.
\item 5.\ Repeat at 2-loop order. 
  \end{description}
\end{flushleft}
\end{minipage}

\subsection{$D$-dimensional Lagrangian and its symmetry breaking}

The considered U(1) gauge theory contains a set of fermion fields $\psi_i$
whose right-handed parts are assigned  ``hypercharges''
$\mathcal{Y}_R{}_i$ and which  interact with the gauge field
$A^\mu$. In $D$ dimensions, the fermionic part of the Lagrangian can
be written as\footnote{%
  For more details and the full form of the $D$-dimensional classical action see Ref.\ \cite{Belusca-Maito:2021lnk}.}
\begin{align}
	\mathcal{L}_\text{fermions} &= i \overline{\psi}_i \slashed{\partial} {\psi}_i + e \mathcal{Y}_R{}_{i} \overline{\psi_R}_i \slashed{A} {\psi_R}_i \, .
\label{Lfermions}  \end{align}
Here $\psi_R=P_R\psi$ with the right-chiral projector
$P_R=(1+\gamma_5)/2$. Note that the kinetic term must involve the
full, $D$-dimensional derivative $\slashed{\partial}$ in order to
generate a regularized, $D$-dimensional propagator denominator in
Feynman diagrams. The mismatch between $\slashed{\partial}{\psi}_i$
and $\slashed{A}{\psi_R}_i$ causes a breaking of gauge invariance in
$D$-dimensions.

On the level of the quantized theory, a gauge fixing
is needed and gauge invariance is replaced by BRST invariance
involving the Faddeev-Popov ghost field $c$, and the
associated symmetry properties of Green functions are expressed by
Ward and Slavnov-Taylor identities. All these can be summarized by the
expression $\mathcal{S}(\Gamma)=0$, where $\mathcal{S}$ is the
Slavnov-Taylor operator and $\Gamma$ the renormalized, finite
generating functional of 1PI Green functions. The breaking of gauge
invariance of the regularized Lagrangian is then equivalent to a
non-zero result of the $D$-dimensional Slavnov-Taylor operator applied to the
classical action $S_0$ in $D$ dimensions,
\begin{align}
\label{DeltahatDef}  \mathcal{S}_d(S_0) &= \widehat{\Delta} \, 
\equiv \int \dInt[d]{x}
        (e \mathcal{Y}_R{}_{i}) c \left\{
            \overline{\psi}_i \left(\overset{\leftarrow}{\widehat{\slashed{\partial}}} \Proj{R} + \overset{\rightarrow}{\widehat{\slashed{\partial}}} \Proj{L}\right) \psi_i
        \right\}
    \, .
\end{align}
Here the quantity $\widehat{\Delta}$, announced in the previous sections,
has been defined. It is an evanescent operator, i.e.\ a local field
operator product which vanishes in 4 dimensions. It  originates
directly from the mismatch between the two terms in
Eq.\ (\ref{Lfermions}). It can be translated into the Feynman rule
used already in the diagram of
Fig.\ \ref{fig:WI}(c). $\widehat{\Delta}$ contains the essence of the
difficulties and provides
the basis of our method of determining the symmetry breaking at the
loop level.

  \subsection{1-loop UV divergences}
As the first step of renormalization of the theory we determine the
1-loop UV divergences by computing the $1/\epsilon$ poles of all
power-counting divergent 1-loop 1PI Green functions. As a result we
obtain the required set of divergent 1-loop counterterms. The
corresponding counterterm action $S_\text{sct}^{1}$ can be decomposed as
  \begin{align}
	S_\text{sct}^{1} = S_{\text{sct,inv}}^{1} + S_{\text{sct,break}}^{1} \, ,
  \end{align}
where the first term $S_{\text{sct,inv}}^{1}$ originates in the
familiar way from field and parameter renormalization transformations
applied to the tree-level action. We suppress its result here. The
second term $S_{\text{sct,break}}^{1} $ is specific to the BMHV scheme
and results from the breaking of gauge and $D$-dimensional Lorentz
invariance. It can be written as
 \begin{align}
\label{SCT1L}  \begin{split}
    S_{\text{sct,break}}^{1} =\;&
        \frac{- e^2}{16 \pi^2 \epsilon} \frac{\text{Tr}(\mathcal{Y}_R^2)}{3} \left( 2 (\overline{S}_{AA} - S_{AA}) + \int \dInt[d]{x} \frac{1}{2} \bar{A}^\mu \widehat{\partial}^2 \bar{A}_\mu \right)
    \, ,
\end{split}
\end{align}
where the symbol $S_{AA}$ denotes the gauge boson kinetic term of the
classical action.

We see that we need specific, non-symmetric divergent counterterms
which cannot be obtained from field and parameter
renormalization. These counterterms are evanescent, i.e.\ the field
operator expressions vanish in 4 dimensions.\footnote{%
  Note that the appearance of specific counterterms for evanescent interactions
  is well known also in the context of dimensional reduction/the FDH
  scheme; for a review and original references see sec.\ 2.3, 2.4 in \cite{Gnendiger:2017pys}.}

\subsection{1-loop symmetry breaking}

We turn to the evaluation of the 1-loop symmetry breaking caused by
the BMHV scheme. Let us first recall the ultimate structure of the
1-loop renormalized generating functional for 1PI Green functions, which is obtained as
$
\Gamma_{\text{renormalized}} = \text{LIM}_{D\to4}\Gamma_{\text{DReg}}$ where LIM denotes $D\to4$ and setting to zero all evanescent terms. The decisive object $\Gamma_{\text{DReg}}$ is obtained as
\begin{align}
  \Gamma_\text{DReg}^{(1)} &= \Gamma^{(1)} +
  S_{\text{sct}}^{1} + S_{\text{fct}}^{1} \, .
\label{structure1L}\end{align}
It is a sum of the generating functional for regularized 1-loop 1PI Green
functions, $\Gamma^{(1)}$, and the complete 1-loop counterterm action,
which in turn is decomposed into the singular counterterms described
above and the finite counterterms. This finite counterterm action,
$S_{\text{fct}}^{(1)}$, contains the symmetry-restoring counterterms
and is the ultimate output of the computation that follows.

It is determined by the requirement that the renormalized theory
satisfies the Slavnov-Taylor identity, $  {\cal S}_d
(\Gamma_\text{DReg}^{(1)}) =0$ in the limit $D\to4$. If we evaluate
  the Slavnov-Taylor operator on the l.h.s.\ at 1-loop order, the
  divergent parts automatically cancel, and the finite parts can be
  rewritten as
       \begin{align}
\label{STI1L}{  {\cal S}_d (\Gamma_\text{DReg}^{(1)})} &=
{\cal S}_d (\Gamma^{(1)})|_{\text{finite}}
{ + {\cal S}_d S_{\text{fct}}^{1}}\,.
\end{align}
Here the first term on the r.h.s.\ corresponds to the Slavnov-Taylor
operator applied to the regularized 1-loop Green functions --- it
directly generalizes the Ward identity and the two Feynman diagrams of
Fig.\ \ref{fig:WI}(a) discussed above. The second term will be discussed in
the subsequent subsection.

This first term can now be simplified by using the regularized quantum
action principle established in Ref.\ \cite{Breitenlohner:1975hg}, as
       \begin{align}
  \label{eq:QAPforSTI1l}
    {\cal S}_d (\Gamma^{(1)}) = \widehat{\Delta} \cdot \Gamma^{(1)} \, .
     \end{align}
where $\widehat{\Delta}\cdot\Gamma^{(1)}$ denotes 1-loop
regularized 1PI Green functions with one insertion of the vertex
corresponding to the operator $\widehat{\Delta}$ defined in
Eq.\ (\ref{DeltahatDef}) --- this directly generalizes the diagram of
Fig.\ \ref{fig:WI}(c) discussed above.

\begin{figure}[h!]
       \begin{minipage}{\textwidth}
	\centering
	\includegraphics[scale=0.5]{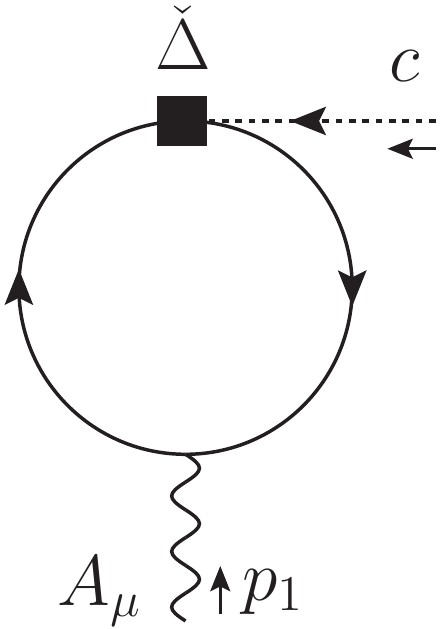}
	\hfill
	\raisebox{-10pt}{\includegraphics[scale=0.5]{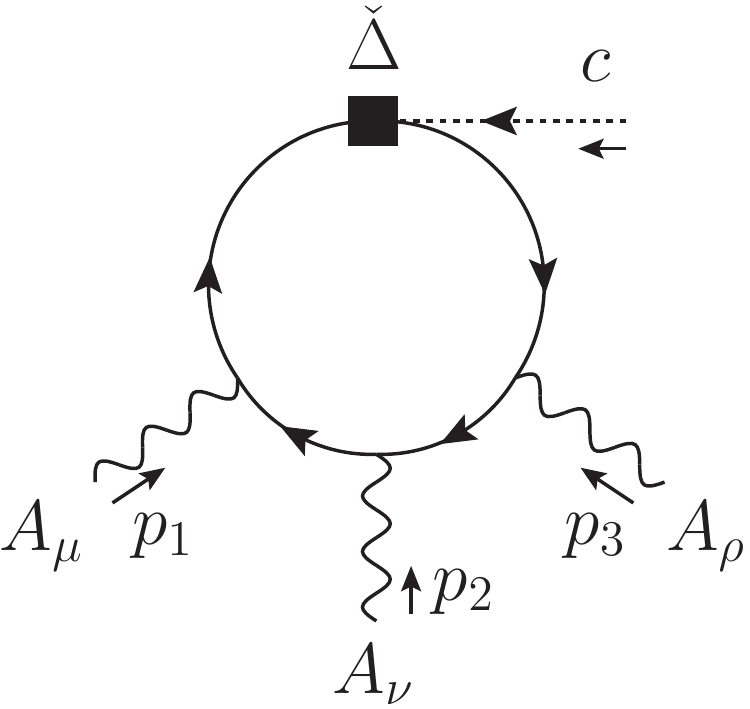}}
	\hfill
	\includegraphics[scale=0.5]{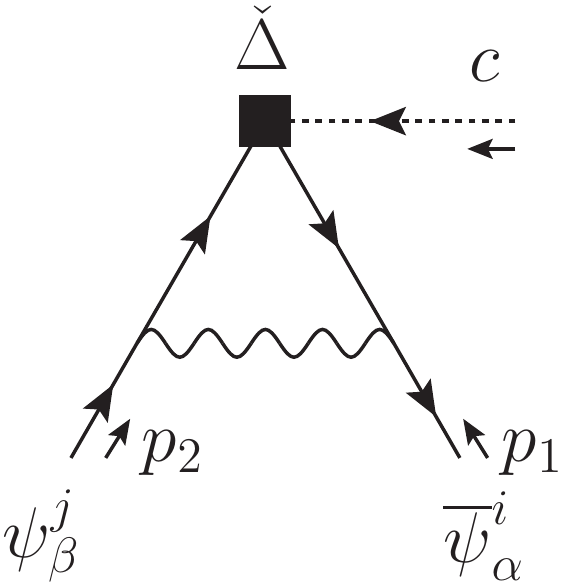}
\end{minipage}
\caption{\label{fig:1L}The three non-vanishing 1-loop diagrams with an insertion of $\widehat{\Delta}$.}
\end{figure}

As a result of this discussion, the recipe for determining the
complete 1-loop symmetry breaking by the BMHV scheme is to evaluate
all non-vanishing 1-loop diagrams with one insertion of
$\widehat{\Delta}$. Since $\widehat{\Delta}$ is evanescent, only
power-counting divergent diagrams can in principle provide a
non-vanishing result in the limit $D\to4$. There are not many such
1-loop diagrams --- in fact, there are precisely four of them. Three
of them are shown in Fig.\ \ref{fig:1L}, the fourth one vanishes provided
the anomaly cancellation condition $	\text{Tr}(\mathcal{Y}_R^3) = 0 $ holds.

Each of these diagrams can be easily evaluated, using the Feynman rule
for the ghost--fermion--fermion vertex corresponding to the
$\widehat{\Delta}$ insertion. In this way, Eq.\ (\ref{eq:QAPforSTI1l}) and thus the
first term on the r.h.s.\ of Eq.\ (\ref{STI1L}) is evaluated.

The interpretation of the three diagrams is obvious. Each of them
describes the violation of a well-known QED-like Ward identity. The
first diagram describes the violation of the transversality of the
photon self energy, the second diagram the violation of the analogous
Ward identity of the photon 4-point function. The third diagram
describes the violation of the Ward identity discussed in section
\ref{nutshell} between the fermion self energy and fermion--gauge boson
three-point function.

\subsection{1-loop symmetry-restoring counterterms}

Our task is now to determine 1-loop symmetry-restoring counterterms,
i.e.\ a counterterm action $S_{\text{fct}}^{(1)}$ which is chosen such
that the sum on the r.h.s.\ of Eq.\ (\ref{STI1L}) vanishes for
$D\to4$, i.e.\
  \begin{align}
{  {\cal S}_d (\Gamma^{(1)})|_{\text{finite}}
 + {\cal S}_d S_{\text{fct}}^{1}} {\stackrel{\text{$D\to4$}}{=}0} \,.
  \end{align}
The expression $ {\cal S}_d S_{\text{fct}}^{1}$ denotes the linearized Slavnov-Taylor operator
applied to the counterterm action, which is essentially the
BRST variation of the counterterm action.

Hence we need to find a counterterm action 
$S_{\text{fct}}^{(1)}$ whose BRST variation is the negative of the
result of the three diagrams of Fig.\ \ref{fig:1L}. This is a
straightforward algebraic exercise. The result is
  \begin{align}
\begin{split}
	S^1_\text{fct} =
	\frac{e^2}{16\pi^2} \int \dInt[4]{x} &\Bigg\{
	\frac{- \text{Tr}(\mathcal{Y}_R^2)}{6} \bar{A}_\mu \overline{\partial}^2 \bar{A}^\mu
	+ \frac{e^2 \text{Tr}(\mathcal{Y}_R^4)}{12} (\bar{A}^2)^2
	\\
&	+ \left(  \frac{5+\xi}{6} \right) \sum_{j}(\mathcal{Y}_R^j)^2 			\Big( \bar{\psi}_j \dot\imath \bar{\slashed{\partial}} \,\Proj{R}\, \psi_j \Big)
\Bigg\}
	\, .
\end{split}
\end{align}
This is the full 1-loop result of the symmetry-restoring
  counterterms for the considered chiral abelian gauge theory in the
  BMHV scheme. We point out:
  \begin{itemize}
  \item
    The result has a very simple structure, contains only three
    terms and can be easily  implemented as a set of additional
    Feynman rules.
  \item
    Each term has an obvious interpretation as a correction term to
    the gauge boson self energy, the quartic gauge boson interaction,
    and to the fermion self energy. The terms are chosen such that they
    guarantee the validity of the three corresponding Ward
    identities. The result contains and generalizes Eq.\ (\ref{eq:FiniteCT1Loop}).
  \item
    The terms are finite and purely 4-dimensional (not
    evanescent). Obviously they are not gauge invariant.
  \end{itemize}
  
\subsection{2-loop UV divergences}

The 2-loop computation proceeds with similar steps. Here we focus on
the essential features of the 2-loop results and point out important
new ingredients and difficulties.

When 1-loop counterterms are taken into account, the remaining UV
divergences at the 2-loop level are local and can be cancelled by a
2-loop counterterm action $	S_\text{sct}^{2}$. Like at 1-loop
level, this can be partially  obtained by field and
parameter renormalization, but a remainder exists. The remainder reads
 \begin{align}
  \begin{split}
	S_{\text{sct,break}}^2 =\;&
		-\frac{e^4}{256\pi^4 \epsilon} \frac{\text{Tr}(\mathcal{Y}_R^4)}{3} \left( 2 (\overline{S}_{AA} - S_{AA})
		+ \left( \frac{1}{2\epsilon} - \frac{17}{24} \right)
		\int \dInt[d]{x} \frac{1}{2} \bar{A}^\mu \widehat{\partial}^2 \bar{A}_\mu \right)
		\\
		& - \frac{e^4}{256\pi^4} \sum_{j}\frac{(\mathcal{Y}_R^j)^2}{3\epsilon} \left( \frac{5}{2}(\mathcal{Y}_R^j)^2 - \frac{2}{3} \text{Tr}(\mathcal{Y}_R^2) \right) \overline{S^{j}_{\bar{\psi}\psi_R}}\,.
\end{split}
\end{align}
It contains the same kind of evanescent terms as the 1-loop result
(\ref{SCT1L}), but in addition there is a non-gauge invariant  and
non-evanescent contribution to the fermion self energy $\overline{S^{j}_{\bar{\psi}\psi_R}}$.

\subsection{2-loop symmetry breaking}

In order to determine the 2-loop symmetry breaking and the
symmetry-restoring counterterms we proceed like at the 1-loop level.
We write down the ultimate structure of the 2-loop renormalized
generating functional (including counterterms but before carrying out $\text{LIM}_{D\to4}$), 
\begin{align}
  \Gamma_\text{DReg}^{(2)} &= \Gamma^{(2)} +
  S_{\text{sct}}^{2} + S_{\text{fct}}^{2} \, ,
\end{align}
in terms of its regularized version (including 1-loop counterterms) $\Gamma^{(2)}$
and the 2-loop singular and finite counterterm actions. If this is
inserted into the Slavnov-Taylor identity at the 2-loop level, the
divergent parts automatically cancel, and the following finite
expressions remain,
       \begin{align}
{  {\cal S}_d (\Gamma_\text{DReg}^{(2)}) }&=
{  {\cal S}_d (\Gamma^{(2)})|_{\text{finite}}
 + {\cal S}_d S_{\text{fct}}^{2}}\,.
\end{align}
The first term on the r.h.s.\ can be simpler evaluated using the
quantum action principle,
       \begin{align}
  \label{eq:QAPforSTI}
         {\cal S}_d (\Gamma^{(2)}) = \widehat{\Delta} \cdot \Gamma^{(2)}+\Delta_\text{ct}^1\cdot\Gamma^{(1)}
  \end{align}
where $\Delta_\text{ct}^1$ is defined like $\widehat{\Delta}$ but using
the 1-loop counterterm action. However, the evaluation of this expression is significantly more
involved than at the 1-loop level. 

\begin{figure}[h!]
  \begin{center}
\begin{tabular}{cc}
  \includegraphics[scale=0.4]{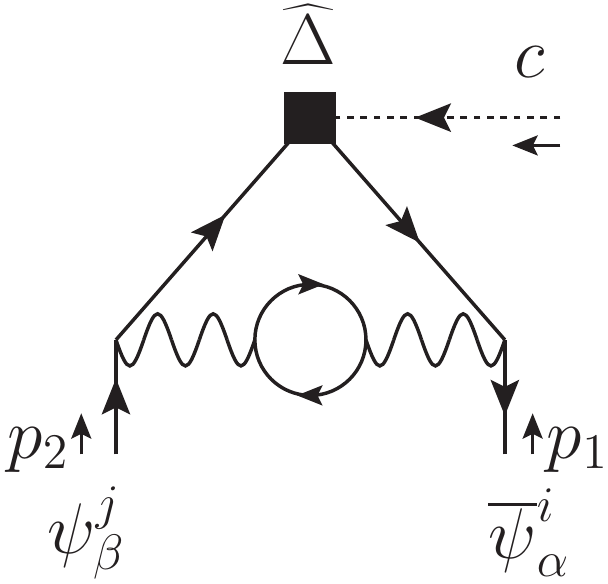}
&      \includegraphics[scale=0.4]{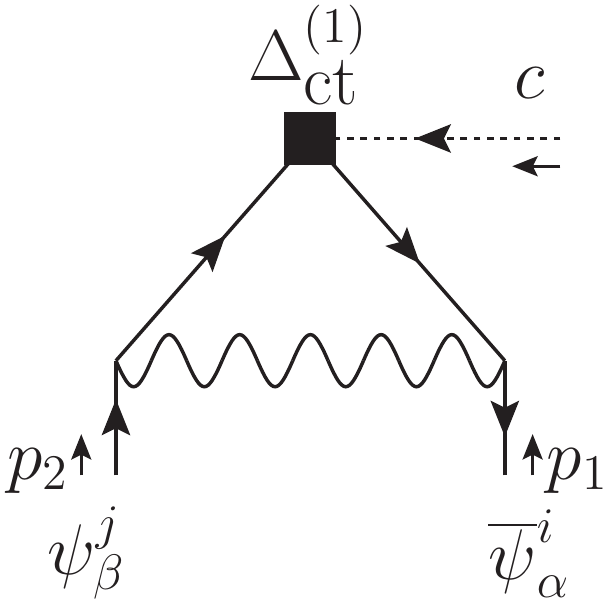}
  \\
  2-loop insertion of $\widehat\Delta$
&   1-loop insertion of $\Delta_\text{ct}^1$
  \\
  \includegraphics[scale=0.4]{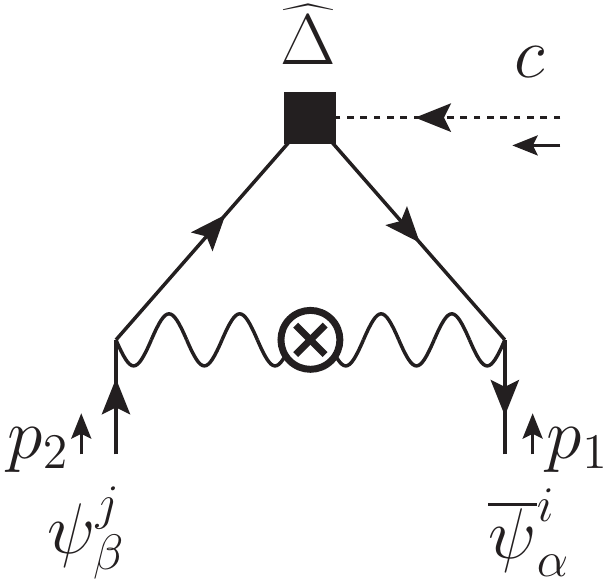}
  &
  \includegraphics[scale=0.4]{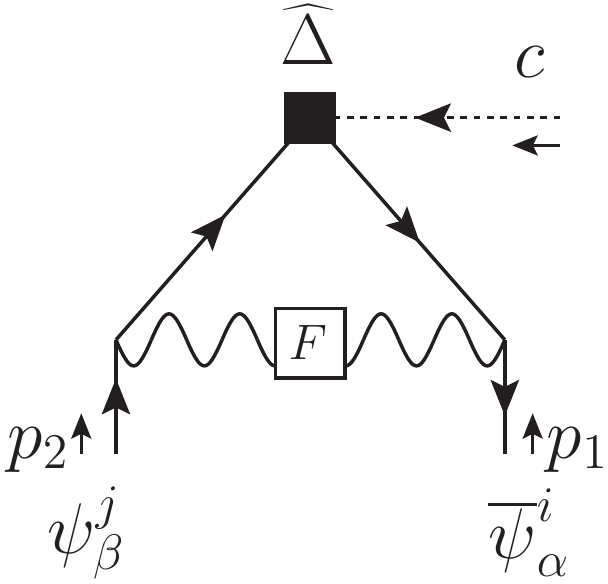}
  \\
\multicolumn{2}{c}{insertion of $\widehat\Delta$ into 1-loop diagram with 1-loop ct insertion}
\end{tabular}
  \end{center}
\caption{\label{fig:2L}The four types of diagrams contributing to
  Eq.\ (\ref{eq:QAPforSTI}).}
\end{figure}

Instead of three non-vanishing diagrams, there are now four different types of
diagrams, each with many concrete examples. They are exemplified in
Fig.\ \ref{fig:2L} with sample diagrams with external ghost--fermion--fermion.
Nevertheless, after evaluating all these diagrams the result acquires
a form like at the 1-loop level. In particular, the cancellation of
non-local terms (non-polynomial in momenta) provides a strong check of
the calculation.

\subsection{2-loop symmetry-restoring counterterms}

Requiring the renormalized Slavnov-Taylor identity to hold at the
2-loop level means
  \begin{align}
{  {\cal S}_d (\Gamma^{(2)})|_{\text{finite}}
 + {\cal S}_d S_{\text{fct}}^{2}}
 {\stackrel{\text{$D\to4$}}{=}0}\,.
  \end{align}
  The first term is computed in terms of the diagrams illustrated by
  Fig.\ \ref{fig:2L}, so this equation determines the desired
  symmetry-restoring finite counterterms $ {\cal S}_d
  S_{\text{fct}}^{2}$. The result is
    \begin{align}
\label{eq:FiniteCT2Loop}
\begin{split}
	S^{2}_\text{fct} =\;
\frac{	e^4 }{	\left(16\pi^2\right)^2} \int \dInt[4]{x}&
\Bigg\{
		 \text{Tr}(\mathcal{Y}_R^4) \frac{11}{48} \bar{A}_\mu \overline{\partial}^2 \bar{A}^\mu
		+ e^2 \frac{\text{Tr}(\mathcal{Y}_R^6)}{8}\, (\bar{A}^2)^2
		\\
		- \sum_{j}(\mathcal{Y}_R^j)^2 \,
		&
			\left( \frac{127}{36} (\mathcal{Y}_R^j)^2 - \frac{1}{27} \text{Tr}(\mathcal{Y}_R^2) \right)
			\Big( \bar{\psi}_j \dot\imath \bar{\slashed{\partial}} \,\Proj{R}\, \psi_j \Big)
		\Bigg\}\,.
\end{split}
\end{align}
This is the full 2-loop result of the symmetry-restoring
  counterterms for the considered chiral abelian gauge theory in the
  BMHV scheme. We point out:
  \begin{itemize}
  \item The result is as simple as at the 1-loop level. Again it can
    be easily  implemented as additional Feynman rules.
  \item
    It contains
    only the same three kinds of terms, corresponding to the gauge boson
    self energy, quartic gauge boson interaction and fermion self
    energy.
  \item
    The only difference are the prefactors, which are now of 2-loop order.
  \end{itemize}

\section{Summary and outlook}

The BMHV scheme of dimensional regularization defines  $\gamma_5$ as a
purely 4-dimensional object which does not fully anticommute in $D$
dimensions. The key advantages of this scheme are the full mathematical
consistency and the existence of all-order proofs of renormalization
properties such as cancellation of divergences and the quantum action
principle. Its disadvantage is the spurious breaking of gauge
invariance in chiral gauge theories.

Here we presented an approach to systematically determine the required
symmetry-restoring counterterms which cancel this breaking of gauge
invariance. The approach is based on evaluating Feynman diagrams with
insertions of the breaking $\widehat{\Delta}$ and is thus simpler than
explicitly evaluating required Ward/Slavnov-Taylor identities (as
illustrated by Fig.\ \ref{fig:WI}). The insertion $\widehat{\Delta}$
corresponds to the tree-level breaking of gauge invariance and thus
encapsulates the core difficulty of the scheme.

We explained the calculational procedure and the structure of
contributing Feynman diagrams up to the 2-loop level. As discussed in
the previous sections, a crucial observation is the simplicity of the
results. It will be straightforward to take into account the obtained
symmetry-restoring counterterms in practical calculations and/or to
implement them in computer-algebra frameworks.

A second important observation is that the structure of the result
does not change between 1-loop and 2-loop order. In general it is
clear that the number of required symmetry-restoring counterterms is
finite (in practice, it is small) and limited by power counting. In
case of the Yang-Mills theory treated in
Ref.\ \cite{Belusca-Maito:2020ala} the set of symmetry-restoring
counterterms comprises all two-point functions and all gauge boson self
interactions, as well as a subset of the interactions between gauge
bosons and matter fields.

There is no obstacle to apply the method to the full electroweak
standard model and to higher loop orders, and work in these directions
is in progress.

\end{document}